\documentclass[12pt]{article}
\usepackage{amsmath,amssymb,amsthm}

\advance\topmargin by -\headheight
\advance\topmargin by -\headsep
\evensidemargin=\oddsidemargin
\advance\hoffset by -3mm
\advance\voffset by  8mm

\makeatletter
\def\section{\@startsection{section}{1}{\z@}{-3.5ex plus -1ex minus
  -.2ex}{2.3ex plus .2ex}{\large\bf}}
\def\subsection{\@startsection{subsection}{2}{\z@}{-3.25ex plus -1ex
  minus -.2ex}{1.5ex plus .2ex}{\normalsize\bf}}
\makeatother

\renewcommand{\a}{\alpha}          
\renewcommand{\b}{\beta}           
\newcommand{\C}{\mathbb{C}}        
\newcommand{\Dl}{\Delta}           
\newcommand{\del}{\partial}        
\newcommand{\diff}{{\mathrm{diff}}} 
\newcommand{\dl}{\delta}           
\newcommand{\eps}{\epsilon}        
\newcommand{\F}{\mathcal{F}}       
\newcommand{\Ga}{\Gamma}           
\newcommand{\ga}{\gamma}           
\renewcommand{\H}{\mathcal{H}}     
\newcommand{\K}{\mathcal{K}}       
\newcommand{\la}{\lambda}          
\newcommand{\loc}{{\mathrm{loc}}}  
\newcommand{\nn}{\nonumber}        
\DeclareMathOperator{\Pf}{Pf}      
\newcommand{\prim}{{\mathrm{prim}}} 
\newcommand{\R}{\mathbb{R}}        
\newcommand{\set}[1]{\{\,#1\,\}}   
\newcommand{\sgn}{{\rm sgn}}       
\renewcommand{\SS}{\mathcal{S}}    
\newcommand{\Sf}{\mathbb{S}}       
\newcommand{\thalf}{\tfrac{1}{2}}  
\newcommand{\tquarter}{\tfrac{1}{4}} 
\def\<#1,#2>{\langle#1,#2\rangle}  

\theoremstyle{plain}
\newtheorem{thm}{Theorem}           
\newtheorem{prop}[thm]{Proposition} 

\theoremstyle{definition}
\newtheorem{defn}{Definition}       

\let\savednewpage=\newpage
\newcommand{\fakenewpage}{\relax}
\newcommand{\suppressnewpage}{\let\newpage=\fakenewpage}
\newcommand{\restorenewpage}{\let\newpage=\savednewpage}

\title{Improved Epstein--Glaser renormalization\\
in coordinate space I. Euclidean framework}

\author{
Jos\'e M. Gracia-Bond\'{\i}a$^{a,b,c}$
\\[1pc]
$^a\,$Forschungszentrum BiBoS, Fakult\"at der Physik\\
Universit\"at Bielefeld, D--33615 Bielefeld, Germany\\
$^b\,$Departamento de F\'{\i}sica Te\'orica, Universidad de Zaragoza\\
E--50009 Zaragoza, Spain\\
$^c\,$Departamento de F\'{\i}sica Te\'orica I, Universidad Complutense\\
E--28040 Madrid, Spain
\\[1pc]
}

\begin{document}

\begin{flushright}
BiBoS/01--12--069\\
DFTUZ/02/01\\
FT--UCM/20--2002\\
January 2002
\end{flushright}

\suppressnewpage  
\maketitle
\restorenewpage   

\vskip 1.5cm

\begin{abstract}
In a series of papers, we investigate the reformulation of
Epstein--Glaser renormalization in coordinate space, both in analytic
and (Hopf) algebraic terms. This first article deals with analytical
aspects. Some of the (historically good) reasons for the divorces of
the Epstein--Glaser method, both from mainstream quantum field theory
and the mathematical literature on distributions, are made plain; and
overcome.
\end{abstract}

\section{Introduction}
\label{sec:introibo}

This is the first of a series of papers, the
companions~\cite{Bettina,Flora} often being denoted respectively II
and~III.
\vfill\eject
We find it convenient to summarize here the aims of these papers, in
reverse order. Ever since Kreimer perceived a Hopf algebra lurking
behind the forest formula~\cite{DirkOriginal}, the question of
encoding the systematics of renormalization in such a structure (and
the practical advantages therein) has been in the forefront. Connes
and Kreimer were able to show, using the $\varphi^3_6$ model as an
example, that renormalization of quantum field theories in momentum
space is encoded in a commutative Hopf algebra \textit{of Feynman
graphs} $\H$, and the Riemann--Hilbert problem with values in the
group of loops on the dual of $\H$~\cite{CKI,CKII}. The latter makes
sense only in the context of renormalization by dimensional
regularization~\cite{Orange,ArgentinaBrasiu}, physicists' method of
choice. Now, whereas it is plausible that the Hopf algebra approach to
renormalization is consistent with all main renormalization methods,
there is much to be learned by a systematic verification of this
conjecture. Paper III focuses on combinatorial-geometrical aspects of
this approach to perturbative renormalization in QFT within the
framework of the Epstein--Glaser (EG) procedure~\cite{EG}.

One can argue that all that experiments have established is (striking)
agreement with (renormalized) momentum space
integrals~\cite{Martinus}. Be that as it may, renormalization on real
space is more intuitive, in that momentum space formulations ``rather
obscure the fact that UV divergences arise from purely short-distance
phenomena''~\cite{Collins}. For the questions of whether and how
configuration space-based methods exhibit the Hopf algebraic
structure, the EG method was a natural candidate. It enjoys privileged
rapports with external field
theory~\cite{HeidelbergTrier,Bellissard,Halley}, possesses a stark
reputation for rigour, and does not share some limitations of
dimensional regularization ---allowing for renormalization in curved
backgrounds~\cite{BrFr}, for instance.

In spite of its attractive features, EG renormalization still remains
outside the mainstream of QFT. The (rather rigorous) QFT text by
Itzykson and Zuber has only the following to say about it: ``\dots the
most orthodox procedure of Epstein and Glaser relies directly on the
axioms of local field theory in configuration space. It is free of
mathematically undefined quantities, but hides the multiplicative
structure of renormalization''~\cite[p.~374]{IZ}. Raymond Stora, today
the chief propagandist of the method, had commented: ``In spite of its
elegancy and accuracy this theory suffers from one defect, namely it
does not yield explicit formulae of actual computational
value''~\cite{LagrSt}. Indeed.

Over the years, some of the awkwardness of the original formalism was
dispelled in the work by Stora. The ``splitting of distributions'' was
reformulated in~\cite{PoSt} as a typical problem of extension (through
the boundaries of open sets) in distribution theory. Moreover,
in~\cite{EllipticSt} it was made clear that an (easier) Euclidean
analog of the EG construction does exist. Beyond being interesting on
its own right (for instance for the renormalization group approach to
criticality), it allows performing EG renormalization in practice by a
(sort of) ``Wick rotation'' trick ---the subject of paper~II of this
series.

\smallskip

When tackling the compatibility question of EG renormalization and the
Connes--Kreimer algebra, two main surviving difficulties are brought
to light. The first is that, while the Hopf algebra elucidation of
Bogoliubov's recursive procedure is defined graph-by-graph, in the EG
approach it is buried under operator aspects of the time-ordered
products and the $\Sf$-matrix, not directly relevant for that
question. This problem was recently addressed by
Pinter~\cite{Gudrunpaper,Gudrun} and also in~\cite{Etoile}; the last
paper, however, contains a flaw, examined in~III.

The second difficulty, uncovered in the course of the same
investigation, has to do with prior, \textit{analytical} aspects of
Epstein and Glaser's basic method of subtraction. For, it was curious
to observe, the extension method by Epstein and Glaser has remained
\textit{divorced as well} from the literature on distributions,
centered mainly on analytical continuation and ``finite part''
techniques. One scours in vain for any factual link between EG
subtraction and the household names of mathematical distribution
theory.

And so the vision of casting all of quantum field theory in the light
of distribution analysis~\cite{Guetti,BP} has remained unfulfilled.

In the present paper we are concerned with the second of the mentioned
difficulties. This means in practice that we deal with primitively UV
divergent diagrams. (Nonprimitive diagrams are dealt with in~III.) By
means of a seemingly minute departure from the letter, if not the
spirit, of the EG original prescription we succeed to deliver its
\textit{missing link} to the standard literature on extension of
distributions. Then we proceed to show the \textit{dominant place} our
improved subtraction method occupies with regard to dimensional
regularization in configuration space; differential
renormalization~\cite{FJL}; ``natural'' renormalization~\cite{NR}; and
BPHZ renormalization.

The benefits of the improved prescription do not stop there: it goes
on to remarkably simplify the task of constructing covariant
renormalizations in~II, and the Hopf-geometrical constructions in~III.

An important sideline of this paper is the use of the theory of
Ces\`aro summability of distributions~\cite{Odysseus,CesarRicardo} in
dealing with the infrared difficulties; this helps to clarify the
logical dependence of the BPHZ procedure on the causal one, already
pointed out in~\cite{PrangeI}. Improved BPHZ methods for massless
fields ensue as well.

The main theoretical development is found in Section~2. Afterwards, we
proceed by way of alternating discussions and examples. In order to
deliver the argument without extraneous complications, we work out
diagrams belonging to scalar theories. Most examples are drawn from
the massless $\varphi^4_4$ model: masslessness is more challenging and
instructive, because of the attendant infrared problems, and more
interesting for the renormalization group calculations performed
in~III. Eventually we bring in examples in massive theories as well.

\section{Renormalization in configuration space}
\label{sec:base-diag}

\subsection{The new prescription}
\label{sec:real-thing}

All derivatives in this paper, unless explicitly stated otherwise, are
in the sense of distributions. We tacitly use the translation
invariance of Feynman propagators and amplitudes; in particular, the
origin stands for the main diagonal.

Let $d$ denote the dimension of the coordinate space. Typically, $d$
will be $4n$. An unrenormalized Feynman amplitude $f(\Ga)$, or simply
$f$, associated to a graph $\Ga$, is smooth away from the diagonals.
We say that $\Ga$ is \textit{primitively} divergent when $f(\Ga)$ is
not locally integrable, but is integrable away from zero. Denote by
$\F_\prim(\R^d) \hookrightarrow L_\loc^1(\R^d\setminus\{0\})$ this
class of amplitudes.

By definition, a tempered distribution $\tilde{f}\in \SS'(\R^d)$ is
an \textit{extension} or \textit{renormalization} of~$f$ if
$$
\tilde{f}[\phi] := \langle\tilde{f},\phi\rangle =
\int_{\R^d} f(x)\phi(x) \,d^d x
$$
holds whenever $\phi$ belongs to $\SS(\R^d\setminus\{0\})$.

Let
\begin{equation}
f(x) = O(|x|^{-a}) \qquad\mbox{as } x \to 0,
\label{eq:alg-sing}
\end{equation}
with $a$ an integer, and let $k = a - d \geq 0$. Then,
$f \notin L_\loc^1(\R^d)$. But $f$ can be regarded as a well-defined
functional on the space $\SS_{k+1}(\R^d)$ of Schwartz functions that
vanish at the origin at order $k+1$. Thus the simplest way to get an
extension of $f$ would appear to be standard Taylor series surgery: to
throw away the $k$-jet of $\phi$ at the origin, in order to define
$\tilde f$ by transposition. Denote this jet by $j^k_0\phi$ and the
corresponding Taylor remainder by $R^k_0\phi$. We have by that
definition
\begin{equation}
\<\tilde f, \phi> = \<f, R^k_0\phi>.
\label{eq:bas-def}
\end{equation}
Using Lagrange's integral formula for the remainder:
$$
R^k_0\phi(x) = (k+1) \sum_{|\b|=k+1}\frac{x^\b}{\b!}
\int_0^1 dt\, (1-t)^k\,\del^\b\phi(tx),
$$
where we have embraced the usual multiindex notation, and exchanging
integrations, one appears to obtain an explicit integral formula for
$\tilde f$:
\begin{equation}
\tilde f(x) = (-)^{k+1} (k+1) \sum_{|\b|=k+1}
\del^\b \biggl[\frac{x^\b}{\b!} \int_0^1 dt\,\frac{(1-t)^k}{t^{k+d+1}}
f\Bigl(\frac{x}{t}\Bigr) \biggr].
\label{eq:bas-Lagr}
\end{equation}

Lest the reader be worried with the precise meaning
of~\eqref{eq:alg-sing}, we recall that in QFT one usually considers a
generalized homogeneity degree, the \textit{scaling degree}~\cite{OS}.
The scaling degree $\sigma$ of a scalar distribution $f$ at the origin
of $\R^d$ is defined to be
$$
\sigma_f = \inf \set{s : \lim_{\la\to 0} \la^s f(\la x) = 0}
\qquad\mbox{for } f\in \SS'(\R^d),
$$
where the limit is taken in the sense of distributions. Essentially,
this means that $f(x) = O(|x|^{-\sigma_f})$ as $x \to 0$ in the
Ces\`aro average sense~\cite{Ricardo}. Then $[\sigma_f]$ and
respectively $[\sigma_f] - d$ ---called the \textit{singular order}---
occupy the place of $a$ in~\eqref{eq:alg-sing} and of~$k$.

The trouble with~\eqref{eq:bas-Lagr} is that the remainder is not a
test function, so, unless the infrared behaviour of $f$ is very good,
we end up in~\eqref{eq:bas-def} with an undefined integral. In fact,
in the massless theory $f$ is an homogeneous function with an
algebraic singularity, the infrared behaviour is pretty bad, and $-d$
is also the critical exponent. A way to avoid the infrared problem is
to \textit{weight} the Taylor subtraction. Epstein and
Glaser~\cite{EG} introduced weight functions $w$ with the properties
$w(0) = 1$, $w^{(\a)}(0) = 0$ for $0 < |\a| \leq k$, and projector
maps $\phi \mapsto W_w\phi$ on $\SS(\R^d)$ given by
\begin{equation}
W_w\phi(x) := \phi(x) - w(x)\, j^k_0\phi(x).
\label{eq:bas-EG}
\end{equation}
The previous ordinary Taylor surgery case corresponds to $w \equiv 1$,
and the identity
$$
W_w(w\phi) = w\, W_1\phi
$$
tells us that $W_w$ indeed is a projector, since $W_w(wx^\ga) = 0$ for
$|\ga| \leq k$.

\smallskip

Look again at~\eqref{eq:bas-EG}. There is a considerable amount of
overkilling there. The point is that, in the homogeneous case, a worse
singularity at the origin entails a better behaviour at infinity. So
we can, and should, weight only the \textit{last} term of the Taylor
expansion. This leads to the definition employed in this paper, at
variance with Epstein and Glaser's:
\begin{equation}
T_w\phi(x) :=  \phi(x) - j^{k-1}_0(\phi)(x)
  - w(x) \sum_{|\a|=k} \frac{x^\a}{\a!} \,\phi^{(\a)}(0).
\label{eq:imp-EG}
\end{equation}
Just $w(0) = 1$ is now required in principle from the weight function.

An amazing amount of mathematical mileage stems from this simple
physical observation. To begin with, $T_w$ is also a projector. To
obtain an integral formula for it, start from
$$
T_w\phi = (1 - w) R^{k-1}_0\phi + w R^k_0\phi,
$$
showing that it $T_w$ interpolates between $R^k_0$, guaranteeing a
good UV behaviour, and $R^{k-1}_0$, well behaved enough in the
infrared. By transposition, using~\eqref{eq:bas-Lagr}, we derive
\begin{align}
T_w f(x) &= (-)^k k
\sum_{|\a|=k} \del^\a \biggl[ \frac{x^\a}{\a!}
\int_0^1 dt\, \frac{(1-t)^{k-1}}{t^{k+d}} f\Bigl(\frac{x}{t}\Bigr)
\Bigl(1-w\Bigl(\frac{x}{t}\Bigr)\Bigr)\biggr]
\nn \\
&\qquad + (-)^{k+1} (k+1) \sum_{|\b|=k+1}
\del^\b \biggl[ \frac{x^\b}{\b!} \int_0^1 dt\,
\frac{(1-t)^k}{t^{k+d+1}} f\Bigl(\frac{x}{t}\Bigr)
w\Bigl(\frac{x}{t}\Bigr) \biggr].
\label{eq:Berkeley-jump}
\end{align}
This is the central formula of this paper.

\subsection{On the auxiliary function}
\label{sec:have-leeway}

It is important to realize what is (and is not) required of the weight
function $w$, apart from a good behaviour at the origin: in view of
the smoothness and good properties of $f$ away from the origin, we
have a lot of leeway, and, especially, $w$ does \textit{not} have to
be a test function, nor to possess compact support. Basically, what is
needed is that $w$ decay at infinity in the weak sense that it sport
momenta of sufficiently high order.

We formalize this assertion for greater clarity. First, one says that
the distribution $f$ is of order $|x|^l$ (with $l$ not a negative
integer) at infinity, in the Ces\`aro sense, if there exists a natural
number $N$, and a primitive $f_N$ of $f$ of order $N$, such that $f_N$
is locally integrable for $|x|$~large and the relation
$$
f_N(x) = O(|x|^{N+l})
$$
as $|x| \uparrow\infty$ holds in the ordinary sense. Now, for any real
constant $\ga$, the space $\K_\ga$ is formed by those smooth functions
$\phi$ such that $\del^\a\phi(x) = O(|x|^{\ga - |\a|})$ as
$|x| \uparrow\infty$, for each~$|\a|$. A topology for $\K_\ga$ is
generated by the obvious family of seminorms, and the space $\K$ is
defined as the inductive limit of the spaces $\K_\ga$ as
$\ga \uparrow\infty$. Consider now the dual space $\K'$ of
distributions. The following are
equivalent~\cite{Odysseus,CesarRicardo}:
\vfill\eject
\begin{itemize}
\item
$f \in \K'$.
\item
$f$ satisfies
$$
f(x) = o(|x|^{-\infty})
$$
in the Ces\`aro sense as $|x| \uparrow \infty$.
\item
There exist constants $\mu_\a$ such that
$$
f(\la x) \sim \sum_{\a\geq0} \frac{\mu_\a\,\dl^{(\a)}(x)}{\la^{|\a|+1}}
$$
in the sense of distributions, as $\la \uparrow\infty$.
\item
All the moments $\<f(x), x^\a>$ exist in the sense of Ces\`aro
summability of integrals (they coincide with the aforementioned
constants $\mu_\a$).
\end{itemize}
Any element of $\K'$ which is regular and takes the value 1 at zero
qualifies as a weight ``function''. For instance, one can take for $w$
an exponential function $e^{iqx}$, with $q \neq 0$. This vanishes at
$\infty$ to all orders, in the Ces\`aro sense, and so it is a
perfectly good infrared problem-buster auxiliary function. The fact
that $e^{iqx} \in \K'$, for $q \neq 0$, means that, outside the origin
in momentum space, the Fourier transform of elements $\phi \in \K$ can
be computed by a standard Ces\`aro evaluation
$$
\hat\phi(q) = \<\exp(iqx), \phi(x)>.
$$
Of course, for this auxiliary function the original
equation~\eqref{eq:bas-EG} no longer applies, since it has no
vanishing derivatives at the origin. But~\eqref{eq:bas-EG} can be
replaced by the more general
\begin{equation}
W_w\phi(x) := \phi(x) - w(x) \sum_{0\leq|\a|\leq k} \frac{x^\a}{\a!}
\biggl( \frac{\phi}{w} \biggr)^{(\a)}(0).
\label{eq:bas-EGbis}
\end{equation}
This was seen, at the heuristic level, by~Prange~\cite{PrangeI}; see
the discussion on the BPHZ formalism in subsection~5.3, where the
``Ces\`aro philosophy'' comes into its own.

These observations are all the more pertinent because the contrary
prejudice is still widespread. For instance, the worthy
thesis~\cite{Gudrun}, despite coming on the footsteps
of~\cite{PrangeI}, yet unfortunately exhibits it; on its page~30:
``\dots [the exponential] function is not allowed in the $W$-operation
because it does not have compact support.'' Of course it is allowed:
then the Fourier transformed subtraction $W_w$ of Epstein and Glaser
becomes exactly the standard BPHZ subtraction, around momentum
$p = q$. What $T_w$ becomes will be revealed later.

\subsection{Properties of the $T$-projector}
\label{sec:we-come}

Consider now the functional variation of the renormalized amplitudes
with respect to~$w$. One has
$$
\Bigl< \frac{\dl}{\dl w}T_w f, \psi \Bigr> :=
\frac{d}{d\la} T_{w+\la\psi}\,f\,\biggr|_{\la=0}
$$
for $T_w$, and similarly for $W_w$, by definition of functional
derivative. It is practical to write now
$$
\dl^\a := (-)^{|\a|}\,\frac{\dl^{(\a)}}{\a!},
$$
for this combination is going to appear with alarming frequency.
{}From~\eqref{eq:bas-EG} we would obtain
\begin{equation}
\frac{\dl}{\dl w}W_w f[\cdot] =
- \sum_{|\a|\leq k} f[x^\a\cdot] \,\dl^\a,
\label{eq:funct-two}
\end{equation}
whereas~\eqref{eq:imp-EG} yields
$$
\frac{\dl}{\dl w}T_w f[\cdot] = - \sum_{|\a|=k} f[x^\a\cdot] \,\dl^\a,
$$
independently of $w$ in both cases. Malgrange's theorem says that
different renormalizations of a primitively divergent graph differ by
terms proportional to the delta function and all its derivatives
$\delta^{(\a)}$, up to $|\a| = k$. Thus there is no canonical way to
construct the renormalized amplitudes, the inherent ambiguity being
represented by the undetermined coefficients of the $\delta$'s,
describing how the chosen extension acts on the finite codimension
space of test functions not vanishing in a neighborhood of~$0$.

There is, however, a more ``natural'' way ---in which the ambiguity is
reduced to terms in the higher-order derivatives of $\delta$,
exclusively. This is guaranteed by our choice of $T_w$.

In practice, one works with appropriate 1-parameter (or few-parameter)
families of auxiliary functions, big enough to be flexible, small
enough to be manageable. Recall than in QFT, with $c = \hbar = 1$, the
physical dimension of length is inverse mass. Let then the variable
$\mu$ have the dimension of mass. We consider the change in $T_w f$
when the variable $w$ changes from $w\equiv w(\mu x)$ to
$w((\mu+\dl\mu) x)$, which introduces the Jacobian
$\frac{\dl w}{\dl\mu} = \frac{\del w(\mu x)}{\del\mu}$, yielding
\begin{equation}
\frac{\del}{\del\mu}\,T_{w(\mu x)}f =
- \sum_{|\a|=k} \<f, x^\nu \del^\nu w(\mu x)x^\a> \,\dl^\a.
\label{eq:pre-ren}
\end{equation}
Here we have assumed that $f$ has no previous dependence on~$\mu$.

Enter now the (rotation-invariant) choice
$w_\mu(x) := H(\mu^{-1} - |x|)$, where $H$ is the Heaviside step
function: it not only recommends itself for its simplicity, but it
turns out to play a central theoretical role. The parameter $\mu$
corresponds in our context to 't~Hooft's energy scale in dimensional
regularization ---see subsection~5.1; the limits $\mu \downarrow 0$
and $\mu \uparrow \infty$ correspond to the case $w = 1$ and
respectively to the ``principal value'' of~$f$; in general they will
not exist.

Write $T_\mu f$ for the corresponding renormalizations. With the help
of~\eqref{eq:pre-ren}, one obtains
\begin{equation}
\mu\frac{\del}{\del\mu} T_\mu f =
\sum_{|\a|=k} f\bigl[ \dl(\mu^{-1}-|x|)\,|x|x^\a \bigr]\,\dl^\a.
\label{eq:RG-form}
\end{equation}
For $f$ homogeneous (of order $-d-k$ as it happens), the expression is
actually \textit{independent of} $\mu$, the coefficients of the
$\dl^\a$ being
\begin{equation}
c^\a = \int_{|x|=1} f\,x^\a = \int_{|x|=A} f\,|x| x^\a,
\label{eq:c-coefs}
\end{equation}
with $|\a| = k$ and any $A > 0$. Note that similar extra terms, with
$|\a| < k$, coming out of the formulae~\eqref{eq:funct-two} would
indeed be $\mu$-dependent.

\smallskip

Compute the $T_\mu$ in the massless (homogeneous) case, whereupon one
can pull $f$ out of the integral sign. We get
\begin{align}
T_\mu f(x) &= (-)^k \biggl\{ \sum_{|\a|=k} \del^\a \biggl[
\frac{x^\a f(x)}{\a!} \bigl( 1 - (1-\mu|x|)^k \bigr) \biggr]
\nn \\
&\qquad\qquad + (k+1) \sum_{|\b|=k+1} \del^\b \biggl[
\frac{x^\b f(x)}{\b!} \int_1^{\mu|x|} dt\, \frac{(1-t)^k}{t}
\biggr] \biggr\}.
\label{eq:comp-form}
\end{align}
Formula~\eqref{eq:comp-form} is simpler than it looks: because of our
previous remark on~\eqref{eq:RG-form}, \textit{all} the
$\mu$-polynomial terms in the previous expression for $T_\mu f$ must
cancel. Let us then denote, for $k\geq 1$,
\begin{eqnarray}
H_k:= \sum_{l=1}^k \frac{(-)^{l+1}}{l} \binom{k}{l} =
\binom{k}{1} - \frac{1}{2} \binom{k}{2} +\cdots- (-)^k \frac{1}{k}.
\label{eq:harmonic}
\end{eqnarray}
At least for $\mu|x| \leq 1$, the expression for $T_\mu f$ becomes
$$
T_\mu f(x) = (-)^k (k+1) \sum_{|\b|=k+1}
\del^\b \biggl[ \frac{x^\b f(x)}{\b!} (\log\mu|x| + H_k) \biggr].
$$

By performing the derivative with respect to $\log\mu$ directly on
this formula, one obtains in the bargain interesting formulae for
distribution theory. Namely, for any $f$ homogeneous of degree $-d-k$:
\begin{equation}
(-)^k (k+1) \sum_{|\b|=k+1} \del^\b \biggl[ \frac{x^\b f(x)}{\b!}
\biggr] = \sum_{|\a|=k}\biggl( \int_{|x|=1} f x^\a \biggr) \dl^\a(x).
\label{eq:biscomp-form}
\end{equation}
Thus the final result is
\begin{equation}
T_\mu f(x) = (-)^k (k+1) \sum_{|\b|=k+1} \del^\b \biggl[
\frac{x^\b f(x)}{\b!} \log\mu|x| \biggr] +
H_k \sum_{|\a|=k} c^\a\,\dl^\a(x),
\label{eq:ast-form}
\end{equation}
with the $c^\a$ given by~\eqref{eq:c-coefs}. The resulting expression
is actually valid for all $x$: away from the origin it reduces to
$f(x)$, as it should. It is in the spirit of differential
renormalization, since $f$ is renormalized as the distributional
derivative of a regular object that coincides with $f$ away from the
singularity.

Let us put equation~\eqref{eq:ast-form} to work at once. When
performing multiplicative renormalization in the causal
theory~\cite{Flora}, the relevant property of a renormalized amplitude
turns out to be its \textit{dilatation} or scaling behaviour. This is
not surprising in view of the form of our integral
equation~\eqref{eq:Berkeley-jump}. Now, a further consequence of the
choice of operator $T_w$ is that it modifies the original homogeneity
in a minimal way. Had we stuck to $W_w$
(using~\eqref{eq:biscomp-form}), the relatively complicated form
$$
W_\mu f(\la\cdot) = \la^{-k-d} \, \biggl( W_\mu f
+ \log\la \sum_{|\a|=k} c^\a \,\dl^\a +
\sum_{|\a|<k} a^\a (\la^{k-|\a|} - 1) \,\dl^\a \biggr),
$$
for some $a^\a$, would ensue; whereas for $T_\mu$
{}from~\eqref{eq:ast-form} one obtains
\begin{equation}
[f]_{R,\mu}(\la x) := T_\mu f(\la x) =
\la^{-k-d} \, \biggl([f]_{R,\mu}(x)
+ \log\la \sum_{|\a|=k} c^\a\,\dl^\a(x) \biggr);
\label{eq:rmnj-inhomog}
\end{equation}
or
$$
E[f]_{R,\mu}(x) :=
(-k-d) [f]_{R,\mu}(x) + \sum_{|\a|=k} c^\a\,\dl^\a(x),
$$
where $E$ denotes the Euler operator $\sum_{|\b|=1} x^\b \del^\b$.
 From now on, the notations $[f]_{R,\mu}$ and $T_\mu f$ will be
interchangeably used; when the dependence on $\mu$ need not be
emphasized, we can write $[f]_{R}$ instead.

We invite the reader to prove~\eqref{eq:rmnj-inhomog} directly, by
reworking the argument used for the case $k = 0$
in~\cite[pp.~307--308]{Polaris}.

Let us record here the obvious fact that when we employ two different
prescriptions compatible with our renormalization scheme, the
difference in the results is given by
\begin{equation}
[f]_{R_1} = [f]_{R_2} + \sum_{|\b|=k} C^\b\,\dl^\b,
\label{eq:start-eq}
\end{equation}
for some constants $C^\b$. ``Different prescriptions'' could mean that
we use different weight functions, or different choices of
renormalization conditions, or simply different values of the parameter
$\mu$. In this respect, we note that the Fourier transforms of our
real space renormalized amplitudes (see the end of Section~5) do not
obey standard renormalization conditions on momentum space, so in
particular contexts they might be in need of modification for this
purpose.

\subsection{Supplementary remarks}
\label{sec:we-arrive}

Following~\cite{SmirZav}, with a difference by a factor of 2 in the
definition, we conveniently formalize the first order operator from
the (larger) space $\SS'_1(\R^d)$ of continuous linear functionals on
Schwartz functions that vanish at the origin, appearing in the
foregoing:
$$
S := \sum_{|\b|=1} \del^\b x^\b.
$$
It ``regularizes'' $\SS'_1$, mapping it onto $\SS'$. Clearly, on
tempered distributions $S = d + E$. Therefore $S$ kills any
homogeneous \textit{distribution} of order $-d$, like $\dl$ ---but not
the homogeneous \textit{functionals}! From~\eqref{eq:ast-form}, it is
clear that (the analog in $\xi$-space of) $S$ captures the Wodzicki
\textit{residue} density in the theory of pseudodifferential operators
---see for instance the discussion in Chapter~7 of~\cite{Polaris}.
Analogously, define
$$
S_{k+1} := (k+1)! \sum_{|\b|=k+1} \frac{\del^\b x^\b}{\b!}.
$$
This sends onto $\SS'$ the space of functionals $\SS'_{k+1}(\R^d)$,
dual of the space of Schwartz functions which vanish at the origin up
to order $k+1$. On well-behaved distributions, it is equivalent to
$(S+k)\cdots(S+1)S$: we think of $S_{k+1}$ as an ordered power of
$S_1$, with the coordinate multiplications remaining to the right of
the differential operators. For massless models, our formulae come
close to simply iterating $S$ in this way, as done in~\cite{SmirZav}.

For theoretical purposes, it should be kept in mind that the
$T$-operator remains in the general framework of the Epstein--Glaser
theory: after all, one could always find weight functions $w(\mu)$
such that $W_{w(\mu)}$ is identical to $T_\mu$. In particular, the
singular order of $T_\mu f$ is the same as that of $f$~\cite{BrFr}.
However, we see already that a finer classification needs to be
introduced.

\begin{defn}
A distribution $f = f_1$ is called \textit{associate homogeneous} of
order~1 and degree~$a$ when there exists a \textit{homogeneous}
distribution $f_0$ of degree~$a$ such that
$$
f_1(\la x) = \la^a(f_1(x) + g(\la) \,f_0(x)),
$$
for some function $g(\la)$. It is readily seen that only the logarithm
function can foot the bill for $g$. Then, a distribution $f_n$ is
called \textit{associate homogeneous} of order~$n$ and degree~$a$ when
there exists an associate homogeneous distribution $f_{n-1}$ of order
$n-1$ and degree~$a$ such that
$$
f_n(\la x) = \la^a(f_n(x) + \log\la \,f_{n-1}(x)).
$$
\end{defn}

Clearly the renormalization of primitively divergent graphs in
massless theories, using $T_\mu$, gives rise to associate homogeneous
distributions of order~1. To pass associate homogeneous distributions
of order~1 thru the same machine~\eqref{eq:comp-form}, in order to
obtain a renormalized closed expression, is routine. Assuming no prior
dependence of $f_1$ on $\mu$, one would have to use now
$$
\mu\frac{\del}{\del\mu} T_\mu f_1 = \sum_{|\a|=k}
\biggl( \int f_1\,x^\a  - \log\mu \int f_0\,x^\a \biggr)\,\dl^\a
$$
instead of~\eqref{eq:RG-form}, to simplify the output
of~\eqref{eq:Berkeley-jump}. We omit the straightforward details. Of
course, if $g = [f]_{R;\mu}$ for $f$ primitive, then $T_\mu g = g$;
but diagrams with a renormalized subdivergence provide less trivial
examples. The complete renormalization of massless theories gives rise
exclusively to associate homogeneous distributions~\cite{Flora}.

As remarked in~\cite{ClassicalRussian}, on the finite dimensional
vector space spanned by an homogeneous distribution and its associates
up to a given order, the Euler operator takes the Jordan normal form.
The $S_{k+1}$ operators are nilpotent on that space, when $a = -d-k$.
For example,
$$
S\biggl[\frac{1}{x^4}\biggr]_R  = S\biggl(\frac{1}{x^4}\biggr),
$$
and therefore $S^2\bigl[\frac{1}{x^4}\bigr]_R = 0$. In consonance with
this, equation~\eqref{eq:ast-form} remains valid for $f$ primitively
divergent of the associated homogeneous type.

\smallskip

The main theoretical development of the $T$-operator closes here. We
have found two instances in the QFT literature of the improved causal
method advocated by us~\eqref{eq:imp-EG}: the ``special $R$
operation'' introduced in the outstanding (apparently unpublished)
manuscript~\cite{ThreefromRussia}; and, unwittingly, ``natural''
renormalization~\cite{NR} ---see subsection~5.2. The scaling
properties are discussed in~\cite{ThreefromRussia}; but no explicit
formulae for the renormalized amplitudes are given there. We turn to
matters of illustration and comparison.

\section{Some examples}
\label{sec:Ex-First}

We compute the simplest primitive diagrams relevant for the four-point
function of $\phi^4_4$ theory ---for these logarithmically divergent
graphs, there is of course no difference between $T_w$ and $W_w$. The
following notation will be used in the sequel:
\begin{equation}
\Omega_{d,m} := \int_{|x|=1} x_i^{2m} =
\frac{2\,\Ga(m+\thalf)\pi^{(d-1)/2}}{\Ga(m+\thalf d)},
\label{eq:Om-eq}
\end{equation}
where $i$ labels any component. In particular,
$\Omega_{d,0} =: \Omega_d$, the area of the sphere in dimension~$d$.
The quotients of the $\Omega_{d,m}$ are rational:
$$
\frac{\Omega_{d,m}}{\Omega_d} =
\frac{(2m-1)!!}{(2m+d-2)(2m+d-4)\cdots d}.
$$
The ``propagator'', given by the formula
$$
D_F(x) = \frac{|x|^{2-d}}{(d-2)\Omega_d},
$$
when $d \neq 2$, and by $D_F(x) = (\log|x|)/\Omega_2$ when $d = 2$, is
simply the Green function for the Laplace equation:
$$
\Dl D_F(x) = -\dl(x).
$$
Consider the ``fish'' diagram in $\varphi^4_4$ theory, giving the
first correction to the four-point function. The corresponding
amplitude is proportional to $x^{-4}$. On using~\eqref{eq:ast-form}
with $k = 0$,
$$
\biggl[ \frac{1}{x^4} \biggr]_R =
\del^\b \biggl[ \frac{x^\b\log(\mu|x|)}{x^4} \biggr]
= S_1\, \frac{\log(\mu|x|)}{x^4}.
$$
Next we look at $\mu\frac{\del}{\del\mu}\bigl(T_\mu D_F^2\bigr)$. By
direct computation, on the one hand,
$$
\mu\frac{\del}{\del\mu} \biggl[ \frac{1}{x^4} \biggr]_R =
\del^\b \biggl[ \frac{x^\b}{x^4} \biggr] =
- \frac{1}{2} \Dl\biggl( \frac{1}{x^2} \biggr) = \Omega_4\,\dl(x),
$$
and on the other, according to~\eqref{eq:RG-form}:
$$
\mu\frac{\del}{\del\mu} \biggl[ \frac{1}{x^4} \biggr]_R = c^0\,\dl(x),
\quad\mbox{with residue}\quad
c^0 = \biggl< \frac{1}{x^4}, \dl(\mu^{-1} - |x|)\,|x| \biggr>_{\R^4}
= \Omega_4,
$$
which serves as a check. In this case, as $\mu$ varies from $0$
to~$\infty$, all possible renormalizations of $D_F^2$ are obtained.
One can as well directly check here the equation:
\begin{equation}
T_\mu f(\la x) =
\la^{-4} \bigl[ T_\mu f(x) + \Omega_4 \log\la\,\dl(x) \bigr],
\label{eq:klug-eq}
\end{equation}
for $f(x) = 1/x^4$. Equation~\eqref{eq:klug-eq} contains the single
most important information about the fish graph and is
\textit{essential} for the treatment of diagrams in which it appears
as one of the subdivergences~\cite{Flora}.

Next among the primitive diagrams relevant for the vertex correction
comes the tetrahedron (also called the ``open envelope'') diagram. In
spite of appearances, it is a three-loop graph, as one of the circuits
depends on the others; it is the lowest-order diagram in $\phi^4_4$
theory with the full structure of the four-point function. The
unrenormalized amplitude $f \in \F_\prim(\R^{12})$ is of the form
$$
f(x,y,z) = \frac{1}{x^2 y^2 z^2 (x-y)^2 (y-z)^2 (x-z)^2},
$$
which is logarithmically divergent overall. A funny thing about this
diagram is that the amplitude for it looks exactly the same in
momentum space ---see for instance~\cite{Dirkbook}. Denote by $s$ the
collective variable $(x,y,z)$. Then
$$
[f]_R = S_{(x,y,z)} \biggl[
\frac{\log(\mu|s|)}{x^2 y^2 z^2 (x-y)^2 (y-z)^2 (x-z)^2} \biggr].
$$

Again, the most important information from the diagram concerns its
dilatation properties. Proceeding as above, we obtain
$$
\mu\frac{\del}{\del\mu} [f]_R =: I\,\dl(x),
$$
where, for any $A > 0$,
$$
I = \int_{s^2=A}\, \frac{|s|\,ds}{x^2 y^2 z^2 (x-y)^2 (y-z)^2 (x-z)^2}.
$$
This integral is computable with moderate effort.
First, one rescales variables: $y = |x|u$ and $z = |x|v$, to obtain
$$
I = \int_{\Sf^3} d\biggl( \frac{x}{|x|} \biggr) \int
\frac{d^4u\,d^4v}{u^2v^2(x/|x|-u)^2(u-v)^2(x/|x|-v)^2}.
$$
The calculation is then carried out by means of ultraspherical
polynomial~\cite{AAR, CKTk} techniques. We recall that these
polynomials are defined from
$$
(1 - 2xr + r^2)^{-n} = \sum_{k=0}^\infty C^n_k(x)\,r^n,
$$
for $r < 1$. There follows an expansion for powers of the propagator
$$
\frac{1}{(x-y)^{2n}} = \frac{1}{|y|^n} \sum_{k=0}^\infty C^n_k(xy)
\biggl( \frac{|x|}{|y|} \biggr)^k,
$$
if, for instance, $|x| < |y|$. Using their orthogonality relation
$$
\int_{\Sf^l} d\biggl(\frac{x}{|x|}\biggr)\,
C^n_k\biggl(\frac{xu}{|x||u|}\biggr)
C^n_l\biggl(\frac{xv}{|x||v|}\biggr) =
\dl_{kl} C^n_k\biggl(\frac{uv}{|u||v|}\biggr) \frac{n\,\Omega_d}{k+n},
$$
to perform the angular integrals (in our case $n = 1$, $l = 3$), we
obtain $I$ as the sum of six radial integrals, corresponding to
regions like $|u| < |v| < 1$, and so forth. Each one is equivalent to
$2\pi^6\zeta(3)$. This yields finally the residue $12\pi^6\zeta(3)$
---the geometrical factor $\Omega_4$ is always present. In
consequence, now
$$
T_\mu f(\la x) =
\la^{-12} [T_\mu f(x) + 12\pi^6\zeta(3) \log\la\, \dl(x)].
$$
This is the first diagram which has a nontrivial topology, from the
knot theory viewpoint, and thus the appearance of a $\zeta$-value is
expected~\cite{Dirkbook}.

Consider now the two-loop ``setting sun'' diagram that contributes
to the two-point function in $\varphi^4_4$ theory; it will prove
instructive. One has to renormalize $1/x^6$, and the singular order
is~2. Off~\eqref{eq:ast-form} we read that
\begin{equation}
\biggl[ \frac{1}{x^6} \biggr]_R =
\frac{1}{2} S_3 \biggl[ \frac{\log(\mu|x|)}{x^6} \biggr] +
\frac{3\pi^2}{8}\,\Dl\dl(x).
\label{eq:set-sun}
\end{equation}
Clearly, our formulae come rather close to simply iterating the
operator~$S$, as done in~\cite{SmirZav}. The last term obviously does
not make a difference for the dilatation properties; but we shall soon
strengthen the case for not dropping it. One has
$$
\biggl[ \frac{1}{x^6} \biggr]_R (\la x) =
\la^{-6} \biggl( \Bigl[ \frac{1}{x^6} \Bigr]_R +
\frac{\Omega_4}{8}\log\la\,\Dl\dl(x) \biggr).
$$

The reader may check that using $W_w$ instead of $T_w$ would bring
to~\eqref{eq:set-sun} the extra term $\pi^2\mu^2\dl(x)$, with an
unwelcome $\mu$-power dependence. As we know, this complicates the
dilatation properties for the diagram. The terms polynomially
dependent on $\mu$ are like the ``junk DNA'' of the Epstein--Glaser
formalism, as they carry no useful information on the residues of
QFT~\cite{Flora}.

More generally, for quadratic divergences (such as also appear in the
first (two-vertex) contribution to the two-point function of the
$\varphi^6_3$ and $\varphi^3_6$ theories), one constructs the
extension
\begin{eqnarray}
\bigl[|x|^{-d-2}\bigr]_R = \frac{1}{2} S_3 \biggl(
\frac{\log\mu|x|}{x^{d+2}} \biggr) + \frac{3\Omega_d}{4d} \,\Dl\dl(x),
\label{eq:quadra}
\end{eqnarray}
and
$$
\bigl[|x|^{-d-2}\bigr]_R (\la x) = \la^{-d-2} \biggl(
\bigl[|x|^{-d-2}\bigr]_R + \frac{\Omega_d}{2d} \log\la \,\Dl\dl(x)
\biggr).
$$

\section{The comparison with the mathematical literature}
\label{sec:Ex-Comp}

\subsection{On the real line}

Now we must muster support for the choice of $T_w$ and $T_\mu$. For
the basics of distribution theory, we recommend~\cite{BB}. For
concrete computations, a good place to start is the treatment by
H\"ormander in Section~3.2 of~\cite{Hoermander}, of the extension
problem for the distributions
$$
f(x) = x_+^{-l}, \ x_-^{-l}, \ |x|^{-l}, \ |x|^{-l}\sgn(x), \ x^{-l}
$$
on the real line. Of course, these are not independent: $x_-$ is just
the reflection of $x_+$ with respect to the origin, then
$|x|^{-l} = x_+^{-l} + x_-^{-l}$, and so on; note that $x^{-1}$ is
just the ordinary Cauchy principal value of~$1/x$.

On our side, for instance, $xf(x) = H(x)$ for $f(x) = x_+^{-1}$;
$xf(x) = \sgn(x)$ for $f(x) = |x|^{-1}$; and so on. Then our
formulae~\eqref{eq:ast-form}, for $l$ odd, give
$$
\bigl[x_+^{-l}\bigr]_R =
\frac{(-)^{l-1}}{(l-1)!} \frac{d^l}{dx^l} (H(x) \log(\mu|x|))
+ H_{l-1}\,\dl^{l-1},
$$
or, say,
$$
\bigl[|x|^{-l}\bigr]_R =
\frac{(-)^{l-1}}{(l-1)!} \frac{d^l}{dx^l} (\sgn(x) \log(\mu|x|)
+ 2 H_{l-1}\,\dl^{l-1},
$$
and, for $l$ even, simply
\begin{equation}
\bigl[|x|^{-l}\bigr]_R =
\frac{(-)^{l-1}}{(l-1)!} \frac{d^l}{dx^l} \log|x|.
\label{eq:mag-simpl}
\end{equation}

H\"ormander invokes the natural method of analytic continuation of
$x_+^z$, with $z$ complex, plus residue subtraction at the simple
poles at the negative integers. Our formulae coincide with
H\"ormander's ---see for example his~(3.2.5)--- provided that (a) we
take $\mu = 1$; and (b) $H_k$ defined in \eqref{eq:harmonic} equals
(as anticipated in the notation) the sum of the first $k$ terms of the
harmonic series! This turns out to be the case, although the proof,
that the curious reader can find in~\cite[Ch.~6]{ThreefromTeX}, is not
quite straightforward. Thus we understand that in~\eqref{eq:ast-form}
and similar formulae $H_k$ just means $\sum_{j=1}^k 1/j$.

Encouraged by this indication of being on the right track, we take a
closer look at the analytic continuation method. The point is that the
function $z \mapsto \int_0^\infty x^z \phi(x)\,dx$ for $\Re z > -1$ is
analytic, its differential being
$dz\,\int_0^\infty x^z \log x\,\phi(x)\,dx$. Let us now consider the
analytic continuation definition for $x^z_+$, where for simplicity we
first take $-2 < \Re z < -1$. One gleans
$$
\<x^z_+, \phi> = \int_0^\infty x^z R^0_0\phi(x) \,dx.
$$
We recall the proof of this:
$$
\<x^z_+, \phi> :=
\biggl< \frac{1}{z+1} \frac{d}{dx} x^{z+1}_+, \phi\biggr> =
- \frac{1}{z+1}\,\lim_{\eps\downarrow0} \int_\eps^\infty
x^{z+1} \phi'(x)\,dx.
$$
A simple integration by parts, taking $v = x^{z+1}$ and
$u = \phi(x) - \phi(0)$, completes the argument.

Iterating the procedure, one obtains
$$
\<x^z_+, \phi> = \int_0^\infty x^z R^{l-1}_0\phi(x)\,dx
$$
for $-l-1 < \Re z < -l$, with $l$ a positive integer. At $z = -l$,
however, this formula fails because of the attendant \textit{infrared}
problem. Let us then compute the first two terms of the Laurent
development of~$x^z$: in view of
$$
\<x^z_+, \phi> = \int_0^{\mu^{-1}} x^z R^{l-1}_0\phi(x)\,dx +
\int_{\mu^{-1}}^\infty R^{l-2}_0\phi(x)\,dx
+ \frac{\phi^{(l-1)}(0)\mu^{-(z+l)}}{(l-1)!\,(z+l)},
$$
the pole part is isolated. Therefore
$$
\lim_{z\to-l} \biggl[x_+^z -
\frac{(-)^{l-1}\dl^{(l-1)}(x)}{(l-1)!(z+l)} \biggr] =
T_\mu(x^l_+) - \dl^{l-1}(x) \log\mu.
$$

H\"ormander goes on to consider Hadamard's \textit{finite part}: that
is, for $x_+$, one studies
$$
\int_\eps^\infty x^z \phi(x)\,dx,
$$
where $\phi$ is always a test function, for any $z \in \C$, and
discards the multiples of powers $\eps^{-\theta}$, for nonvanishing
$\theta$ with $\Re\theta \geq 0$, and the multiples of $\log\eps$.
He proves that this finite part coincides with the result of the
analytic continuation method.

We do not need to review his proof, as we can show directly the
identity of our results with finite part, by the following trick:
\begin{align*}
\int_\eps^\infty \frac{\phi(x)}{x^l} \,dx
&= \sum_{j=0}^{l-1} \int_\eps^{\mu^{-1}} \frac{\phi^{(j)}(0)}{j!}
    \,x^{j-l}\,dx + \sum_{j=0}^{l-2} \int_{\mu^{-1}}^\infty
    \frac{\phi^{(j)}(0)}{j!} \,x^{j-l}\,dx
\\
&\qquad + \int_\eps^{\mu^{-1}} x^{-l} R^{l-1}_0\phi(x) \,dx
         + \int_{\mu^{-1}}^\infty x^{-l} R^{l-2}_0\phi(x) \,dx.
\end{align*}
Then, as $\eps \downarrow 0$, the two last terms give rise to the
$T_\mu(1/x^l)$ renormalization and the surviving finite terms cancel,
except for the expected contribution
$-\frac{\phi^{(l-1)}(0)}{(l-1)!} \log\mu$, coming from the first sum.

Denote the finite part of $x_+^{-l}$ by $\Pf \frac{H(x)}{x^l}$, where
$\Pf$ stands for \textit{pseudofunction} (or for \textit{partie
finite}, according to taste). In summary, we have proved:

\begin{prop}
On the real line, the $T$-operator leads to a one-parameter
generalization of the finite part and analytic continuation
extensions, to wit,
$$
\bigl[x_+^{-l}\bigr]_R := T_\mu(x_+^{-l}) =
\Pf \frac{H(x)}{x^l} + \dl^{l-1}(x) \log\mu.
$$
\end{prop}

This generalization is in the nature of things. Actually, the finite
part and analytic continuation methods are not nearly as uniquely
defined as some treatments make them appear. For instance, at the
negative integers the definition of the finite part of $x^z$
\textit{changes} if we substitute $A\eps$ for $\eps$; and,
analogously, one can slip in a dimensionful scale in analytical
prolongation formulae. The added flexibility of the choice of $\mu$ is
convenient.

We parenthetically observe that the nonhomogeneity of $T_\mu$, and
then of $\Pf$, is directly related to the presence of logarithmic
terms in the asymptotic expansion for the heat kernels of elliptic
pseudodifferential operators~\cite{Odysseus}.

Finally, we remark that the Laurent development for $x_+^z|_{z=-l}$
continues:
\begin{equation}
\frac{\phi^{(l-1)}(0)}{\eps(l-1)!} + \Pf(x^{-l}_+) +
\frac{\eps}{2}\Pf(x^{-l}_+ \log x_+) +
\frac{\eps^2}{3!}\Pf(x^{-l}_+ \log^2 x_+) + \cdots
\label{eq:L-key}
\end{equation}
with $\eps := z + l$ and the obvious definition for
$$
\Pf(x^{-l}_+ \log^m x_+) = \bigl[x^{-l}_+ \log^m x_+\bigr]_{R;\mu=1}.
$$

\subsection{Dimensional reduction}

The phrase ``dimensional reduction'' is used in the sense of
ordinary calculus, it does not refer here to the method of
renormalization of the same name. The reader may have wondered why we
spend so much time on elementary distributions on $\R$. The reason, as
it turns out, is that an understanding of the 1-dimensional case is
all that is needed for the renormalization of ${|x|^{-d-k}}$, for any
$k$ and in any dimension $d$; thus covering the basic needs of
Euclidean field theory. For instance, one can define
$\bigl[x^{-4}\bigr]_R$ on $\R^4$ from knowledge of $x^{-1}_+$ on~$\R$.

Denote $r := |x|$ and let $f(r)$ be an amplitude on $\R^d$, depending
only on the radial coordinate, in need of renormalization. We are
ready now to simplify~\eqref{eq:ast-form} by a method that generalizes
Proposition 1 to any number of dimensions.

Given an arbitrary test function $\phi$, consider its projection onto
the radial-sum-values function $\phi \mapsto P\phi$ given by
$$
P\phi(r) := \int_{|y|=1} \phi(ry).
$$
We compute the derivatives of $P\phi$ at the origin:
$(P\phi)^{(2m+1)}(0) = 0$ and
$$
(P\phi)^{(2m)}(0) = \Omega_{d,m}\,\Dl^m\phi(0).
$$
To prove this, whenever all the $\b$'s, and thus $n$, are even, use
$$
(P\phi)^{(n)}(0) =
\sum_{|\b|=n} \frac{n!\,\del^\b\phi}{\b!} \biggr|_{x=0}
\int_{|y|=1} y_1^{\b_1} \dots y_n^{\b_n} =
\frac{2\,\Ga(\frac{\b_1+1}{2})\cdots\Ga(\frac{\b_n+1}{2})}
      {\Ga(\frac{n+d}{2})},
$$
in consonance with~\eqref{eq:Om-eq}; the integral vanishes otherwise.
Note that $P\phi$ can be considered as an even function defined on the
whole real line. Then, whenever the integrals make sense,
$$
\<f(r), \phi(x)>_{\R^d} = \<f(r) r^{d-1}, P\phi(r)>_{\R^+},
$$
which in particular means that extension rules for $H(r)f(r)$ on $\R$
give extension rules for $f(r)r^{d-1}$ on $\R^d$. This we call
dimensional reduction.

Before proceeding, let us put the examined real line extensions in
perspective, by investigating how satisfactory our results are from a
general standpoint, and whether alternative renormalizations with
better properties might exist. Note first, from~\eqref{eq:ast-form}:
$$
\Pf \frac{H(x)}{x} = \frac{d}{dx} (H(x) \log|x|).
$$
For $z$ not a negative integer, the property
\begin{equation}
x\,x_+^z = x_+^{z+1}
\label{eq:req-one}
\end{equation}
obtains; and excluding $z = 0$ as well, we have
\begin{equation}
\frac{d}{dx} x_+^z = z\,x_+^{z-1}.
\label{eq:req-two}
\end{equation}
One can examine how the negative integer power candidates fare in
respect of these two criteria: of course, except for $x^{-l}$, which
keeps all the good properties, homogeneity is irretrievably lost.

Actually, it is $x_+^{-l}$ that we need. One could \textit{define} a
renormalization $\bigl[x_+^{-l}\bigr]_\diff$ of $x_+^{-l}$ simply by
$$
\bigl[x_+^{-l}\bigr]_\diff :=
(-)^{l-1} \frac{1}{(l-1)!} \frac{d^l}{dx^l} (H(x) \log|x|),
$$
so $\bigl[x_+^{-1}\bigr]_\diff = \Pf \frac{H(x)}{x}$; and
automatically the second~\eqref{eq:req-two} of the requirements
$$
\frac{d}{dx} \bigl[x_+^{-l}\bigr]_\diff =
-l\,\bigl[x_+^{-l-1}\bigr]_\diff
$$
would be fulfilled. This would be ``differential renormalization'' in
a nutshell. It differs from the other extensions studied so far: from
our previous results,
$$
\bigl[x_+^{-l}\bigr]_\diff = \bigl[x_+^{-l}\bigr]_R
+ (-)^l (H_{l-1} + \log\mu) \,\dl^{l-1}(x).
$$

On the other hand, it is seen that
$$
\frac{d}{dx} T_\mu(x_+^{-l}) = -l\,T_\mu(x_+^{-l-1}) + \dl^l(x),
$$
so that $T_\mu$ does not fulfill that second requirement; but in
exchange, it does fulfill the first one~\eqref{eq:req-one}:
$$
x_+\,T_\mu(x_+^{-l}) = T_\mu(x_+^{-l+1}).
$$
There is no extension of $x_+^a$ for which both requirements
simultaneously hold.

It looks as if we are faced with a choice between $[\cdot]_\diff$ and
$T_\mu(\cdot)$ ---which is essentially $\Pf(\cdot)$--- each one with
its attractive feature. But the situation is in truth not symmetrical:
in higher dimensional spaces the analogue of the first requirement
\textit{can} be generalized to the renormalization of $|x|^{-l}$;
whereas the analog of the second then \textit{cannot} be made to work
---have a sneak preview at~\eqref{eq:super-machine}.

\smallskip

Estrada and Kanwal \textit{define} then, for $k \geq 0$
\cite{wonderEKI,wonderEKII},
\begin{align*}
\biggl<\Pf\Bigl(\frac{1}{r^{d+k}}\Bigr),  \phi(x)\biggr>_{\R^d} &:=
\biggl<\Pf\Bigl(\frac{1}{r^{k+1}}\Bigr), P\phi(r)\biggr>_{\R^+};
\\[3pt]
\biggl<\Bigl[\frac{1}{r^{d+k}}\Bigr]_\diff,  \phi(x)\biggr>_{\R^d} &:=
\biggl<\Bigl[\frac{1}{r^{k+1}}\Bigr]_\diff, P\phi(r)\biggr>_{\R^+}.
\end{align*}
In view of~\eqref{eq:mag-simpl}, the case $k$~odd is very easy, and
then all the definitions coincide:
$$
\Pf\Bigl(\frac{1}{r^{d+k}}\Bigr) =
T_\mu\Bigl(\frac{1}{r^{d+k}}\Bigr) =
\Bigl[\frac{1}{r^{d+k}}\Bigr]_\diff = r^z\bigr|_{z=-d-k},
$$
the function $r^z$ having a removable singularity at $-d-k$. However,
in most instances in QFT $k$ happens to be even, so we concentrate on
this case. \textit{We} are not in need of new definitions. By going
through the motions of changing to radial plus polar coordinates and
back, one checks that, assuming a spherically symmetric weight
function $w$, the evaluation $\<T_wf(r), \phi(x)>$ is equal to
$$
\biggl<f(r), \phi(x) - \phi(0) - \frac{\Dl\phi(0)}{2!\,d} r^2 -\cdots-
w(r)\frac{\Omega_{d,m}\,\Dl^m\phi(0)}{(2m)!\,\Omega_d}r^{2m}\biggr>;
$$
the right hand side being invariant under $T_w$. This was perhaps
clear from the beginning, from symmetry considerations. It means in
particular that the different putative definitions of $T_\mu$ on
$\R^d$ obtained from $T_\mu$ on the real line all \textit{coincide}
with the original definition, that is:

\begin{prop}
The $T_\mu$ operators are mutually consistent under dimensional
reduction.
\end{prop}

Moreover,
$$
r^{2q}\, T_\mu(r^{-d-2m}) = T_\mu(r^{-d-2m+2q})
$$
follows, by using
the easy identity
$$
r^2 \Dl^m\dl(x) = 2m(2m+d-2) \Dl^{m-1}\dl(x).
$$
Therefore, it is now clear that
$$
T_\mu(r^{-d-2m}) = \Pf(r^{-d-2m})
+ \frac{\Omega_{d,m}\,\Dl^m\dl(x)}{\Omega_d\,(2m)!} \log\mu.
$$

It remains to compute the derivatives. A powerful technique, based on
``truncated regularization'' and calculation of the derivatives across
surface jumps, was developed and clearly explained
in~\cite{wonderEKI}. It is rather obvious that for $k - d$ odd the
``na\"{\i}ve'' derivation formulae (see right below) will apply.
Whereas for $k - d = 2m$ even, they obtain extra delta function terms;
in particular for the powers of the Laplacian
\begin{align}
\Dl^n \biggl[\frac{1}{r^{d+2m}}\biggr]_\diff
&= (d+2m+2n-2)\cdots(d+2m+2)(d+2m)(2m+2)\cdots
\nn \\
&\qquad \times\,(2m+2n) \biggl[\frac{1}{r^{d+2m+2n}}\biggr]_\diff
+ \frac{\Omega_{d,m}}{(2m)!} \sum_{l=1}^n \frac{\Dl^n\dl(x)}{2m+2l-1}.
\label{eq:super-machine}
\end{align}
The first term is what we termed the ``na\"{\i}ve'' formula.

Estrada and Kanwal do not explicitly give the powers of $\Dl$ for
finite part. But from~\eqref{eq:super-machine} is a simple task to
compute
\begin{align*}
\Dl^n \biggl[\frac{1}{r^{d+2m}}\biggr]_R
&= (d+2m+2n-2)\cdots(d+2m+2)(d+2m)(2m+2)\cdots
\\
&\qquad \times\,(2m+2n)\biggl[\frac{1}{r^{d+2m+2n}}\biggr]_R -
\frac{\Omega_{d,m}}{(2m)!} \sum_{l=1}^n
\frac{(4(m+l)+d-2)\,\Dl^n\dl(x)}{2(m+l)(2m+2l+d-2)}.
\end{align*}

No one seems to have computed explicitly the distributional
derivatives of the\\
$\Pf(x^{-l}_+ \log^m x_+)$ and the correspondingly
defined $\Pf(r^{-l} \log^m r)$, although they might be quite helpful
for Euclidean QFT on configuration space.

\smallskip

We next enterprise to tackle a comparison with methods of
renormalization in real space in the physical literature. Of those
there are not many: it needs to be said that the flame-keepers of the
Epstein--Glaser method~\cite{Scharf} actually work in momentum space
(using dispersion relation techniques). Euclidean configuration space
dimensional regularization, on the other hand, starting
from~\cite{CKTk}, evolved into a powerful calculational tool in the
eighties. With the advent of ``differential
renormalization''~\cite{FJL} in the nineties, regularization-free
coordinate space techniques came into their own: they are the natural
``market competitors'' for the ideas presented here.

We deal first with dimensional regularization.

\section{Comparison with the QFT literature}
\label{sec:Ex-Compbis}

\subsection{Dimensional regularization and ``minimal subtraction''}

Dimensional regularization on real space, for primitively divergent
diagrams, can be identified with analytic continuation. To get the
basic idea, it is perhaps convenient to perform first a couple of
blind calculations. Start from the identity
$$
\mu^\eps |x|^{-d+\eps} =
\frac{\mu^\eps}{\eps} S_x\bigl(|x|^{-d+\eps}\bigr).
$$

Then, expanding in $\eps$, on use of~\eqref{eq:ast-form}, it follows
that
$$
\mu^\eps |x|^{-d+\eps} = \Omega_d \frac{\dl(x)}{\eps} +
S_x\,\frac{\log(\mu|x|)}{|x|^d} + O(\eps).
$$
The first term is a typical infinite (as $1/\eps$) counterterm of the
dimensionally regularized theory. The order of the delta function
derivative, 0 in this case, tells us that we are dealing with a
logarithmic divergence. The coefficient $\Omega_n$ of the counterterm,
or QFT residue, coincides with our \textit{scaling coefficient} of
Section~2. The second term is precisely $[1/x^d]_R$, our renormalized
expression.

Let us go to quadratic divergences. A brute-force computation
establishes for them the differential identity
\begin{eqnarray}
\mu^\eps |x|^{-d-2+\eps} =
\frac{\mu^\eps}{2\eps(1-\frac{3}{2}\eps+\frac{1}{2}\eps^2)}
\,S_3 \bigl(|x|^{\eps-d-2}\bigr).
\label{eq:mirac-app}
\end{eqnarray}
On the other hand, from~\eqref{eq:ast-form}:
\begin{eqnarray}
S_3\bigl(|x|^{-d-2}\bigr) = \frac{\Omega_d}{d}\,\Dl\dl(x).
\end{eqnarray}
Performing in~\eqref{eq:mirac-app} the expansion with respect to
$\eps$, this yields
$$
\mu^\eps |x|^{-d-2+\eps} = \frac{\Omega_d}{2d\,\eps} \Dl\dl(x) +
\frac{1}{2}\, S_3\biggl(\frac{\log\mu|x|}{x^{d+2}}\biggr) +
\frac{3\Omega_d}{4d}\,\Dl\dl(x) + O(\eps).
$$
That is,
$$
\mu^\eps |x|^{-d-2+\eps} = \frac{\Omega_d}{2d} \frac{\Dl\dl(x)}{\eps}
+ \bigl[|x|^{d-2}\bigr]_R + O(\eps).
$$
A pattern has emerged: as before, there is a unique counterterm in
$\frac{1}{\eps}$; the residue coincides with our scaling coefficient;
the order of the delta function derivative reminds us of the order of
the divergence we are dealing with; and the ``constant'' regular term
is precisely $[1/|x|^{d+2}]_R$ constructed in~\eqref{eq:quadra}
according to our renormalization scheme.

The correspondence between the two schemes, at the present level, is
absolute and straightforward. It is then a foregone conclusion that we
shall have $\mu$-independent residues, always coincident with the
scaling factors, for the simple poles of $1/|x|^{d+2m}$, and that the
first finite term shall coincide with $T_\mu$, provided we identify
our scale with 't~Hooft's universal one. This is an immediate
consequence of the Laurent development~\eqref{eq:L-key}, transported
to $\R^d$ by dimensional reduction. In symbols
\begin{equation}
\mu^\eps |x|^{\eps-d-2m} =
\frac{\Omega_{d,m}}{(2m)!\,\eps}\,\Dl^m\dl(x)
+ \bigl[|x|^{-d-2}\bigr]_R + O(\eps).
\label{eq:dim-par}
\end{equation}
This substantiates the claim that $T_\mu$ effects a kind of minimal
subtraction. Let us point out, in the same vein, that already
in~\cite{Odysseus} the analytic continuation of Riemann's zeta
function was evaluated as the outcome of a quantum field
theory-flavoured renormalization process.

A word of warning is perhaps in order here. Performing the Fourier
transform of these identities, we do not quite obtain the usual
formulae for dimensional regularization in momentum space. The
nonresemblance is superficial, though, and related to choices of
``renormalization prescriptions''. The beautiful correspondence is
``spoiled'' (modified) as well for diagrams with subdivergences,
because in dimensional regularization contributions will come to
$O(\eps^0)$ from the higher terms of the $\eps$-expansion, when
multiplied by the unavoidable singular factors; but, again, the
difference is not deep: we show in III how one organizes the Laurent
expansions with respect to $d$ so as to make the correspondence with
the $T$-subtraction transparent.

Much was made in~\cite{DirkResidue}, and rightly so, of the importance
of the perturbative residues in the dimensional regularization scheme.
Residues for primitive diagrams are the single most informative item
in QFT. The coefficients of higher order poles are determined by the
residues ---consult the discussion in~\cite{Gross}. Now, the appeal of
working exclusively with well-defined quantities, as we do, would be
much diminished if that information were to disappear in our approach.
But we know it is not lost: it is stored in the scaling properties.

\subsection{Differential renormalization and
  ``natural renormalization'' in QFT}

Differential renormalization, in its original form, turns around the
following extension of $1/x^4$ (in $\R^4$):
\begin{equation}
[1/x^4]_{\rm R,FJL} := -\frac{1}{4}\,\Dl\,\frac{\log\mu^2x^2}{x^2}.
\label{eq:micro-machine}
\end{equation}
At present, two main schools of differential renormalization seem
still in vogue: the original and more popular ``(constrained)
differential renormalization'' of the Spanish school ---see for
instance~\cite{Cetreria}--- and the ``Russian school'' ---inaugurated
in~\cite{SmirZav}. This second method, as already reported, reduces to
systematic use of the operators $S_{k+1}$, i.e., to our
formulae~\eqref{eq:ast-form} without the delta terms. Whereas the
first school has its forte in concrete 1-loop calculations for
realistic theories, assuming compatibility of differentiation with
renormalization, the second initially stressed the development of
global renormalization formulae for diagrams with subdivergences, and
the compatibility of Bogoliubov's rules with renormalization.

Hereafter, we refer mainly to the original version. It proceeded from
its mentioned starting point to the computation of more complicated
diagrams by reductions to two-vertex diagrams. This involves a
bewildering series of tricks, witness more of the ingenuity of the
inventors than of the soundness of the method. V.~gr., the tetrahedron
diagram (considered already) is rather inelegantly renormalized by the
substitution $1/x^2 \mapsto x^2[1/x^4]_{\rm R,FJL}$. They get away
with it, in that particular case, because their expression is still
not infrared divergent. But in nonprimitive diagrams infrared
infinities may arise in relation with the need to integrate the
product of propagators over the coordinates of the \textit{internal}
vertices in the diagram, and, in general, under the procedures of
differential renormalization it is impossible to avoid incurring
infrared problems~\cite{SmirIR}.

Even for primitively divergent diagrams, differential renormalization
is not free of trouble. In his extremely interesting paper,
Schnetz~\cite{NR} delivers a critique of differential renormalization.
In elementary fashion, notice that
$$
\frac{x^\mu\log(\mu^2x^2)}{x^4} =
-\frac{1}{2}\,\del^\mu \biggl[\frac{1+\log(\mu^2x^2)}{x^2}\biggr],
$$
and so
$$
\bigl[1/x^4\bigr]_R = -\frac{1}{4} \Dl\,\frac{1+\log(\mu^2x^2)}{x^2}.
$$
This is to say:
\begin{equation}
\bigl[1/x^4\bigr]_R - \bigl[1/x^4\bigr]_{\rm R,FJL} = \pi^2\,\dl(x).
\label{eq:micro-machinebis}
\end{equation}
We contend that ``our'' $[1/x^4]_R$ and not $[1/x^4]_{\rm R,FJL}$ is
the right definition. Of course, one is in principle free to add
certain delta terms to each individual renormalization and proclaim
that to be the ``right'' definition. However, $1/x^4$ on $\R^4$ is
dimensionally reduced to $x^{-1}_+$ on $\R^+$ and because, as already
pointed out, differential renormalization of this distribution is
consistent with $[x^{-1}_+]_R$ for $\mu=1$, the $[\cdot]_{\rm R,FJL}$
definition is \textit{inconsistent} with any of the natural
alternatives we established in the previous subsection. (It would
clearly induce back an extra $\dl$ term in the definition of
$[x^{-1}_+]_R$ on the real line, fully unwelcome in the context.)

In other words, if we want to make use both of sensible rules of
renormalization for the radial integral (namely, including
differential renormalization at this level) and of Freedman, Johnson
and Latorre's formulae, we have to relinquish the standard rules of
calculus. This Schnetz noticed.

Schnetz proposes instead a ``natural renormalization'' procedure on
$\R^4$, boiling down to the rule
\begin{equation}
\Dl^{n+1}\frac{\log(\mu^2x^2)}{x^2} =
- \biggl[\frac{4^{n+1}n!(n+1)!}{x^{2n+4}}\biggr]_R
+ \biggl(8\pi^2 H_n + \frac{4\pi^2}{n+1}\biggr) \Dl^n\dl(x),
\label{eq:nat-ren}
\end{equation}
whose first instance is precisely the previous
equation~\eqref{eq:micro-machinebis}. This he found by heuristically
defining ``natural renormalization'' as the one that relates
renormalization scales at different dimensions without changing the
definition of ordinary integrals or generating $r$-dependence in the
renormalization of $r$-independent integrals; and by elaborate
computations to get rid of the angular integrals.

His calculation is any rate correct, and the results can be read off
(for $d = 4$, $m = 0$) our~\eqref{eq:super-machine}, taking into
account~\eqref{eq:micro-machine} and ~\eqref{eq:micro-machinebis}. We
have proved that our operator $T_w$ in the context just amounts to
``natural renormalization''.

Shortly after the inception of the differential renormalization, it
naturally occurred to some people that a definite relation should
exist between it and dimensional regularization. However, because of
the shortcomings of the former, they landed on formulae both messy and
incorrect~\cite{DNuria}. The reader is invited to compare them with
our~\eqref{eq:dim-par}.

The more refined version of differential renormalization
in~\cite{SmirZav}, coincides with our formulae for logarithmic
divergences and eludes the main thrust of Schnetz's critique; however,
we have seen that in general it does not yield the Laurent development
of the dimensionally regularized theory either. On the other hand, it
must be said that the emphasis in~\cite{FJL,NR} in bringing in the
Laplacian instead of the less intuitive albeit more fundamental $S_k$
operators has welcome aspects, not only because of the enhanced
feeling of understanding, but also in that it makes the transition to
momentum space a trivial affair, as soon as the Fourier transform of
the (evidently tempered) distribution $x^{-2} \log(\mu^2x^2)$ is
known.

The trinity of basic definitions in differential renormalization is
then replaced by the identities
\begin{align*}
\biggl[\frac{1}{x^4}\biggr]_R
&= - \frac{1}{2} \,\Dl \,\frac{\log\mu|x|}{x^2} + \pi^2\,\dl(x);
\\
\biggl[\frac{1}{x^6}\biggr]_R
&= - \frac{1}{16} \,\Dl^2\, \frac{\log\mu|x|}{x^2}
    + \frac{5}{8} \pi^2 \Dl\dl(x);
\\
\biggl[\frac{\log\mu|x|}{x^4}\biggr]_R
&= - \frac{1}{4} \,\Dl\, \frac{\log^2\mu|x| + \log\mu'|x|}{x^2}
    + \frac{\pi^2}{2} \,\dl(x);
\end{align*}
the $\dl$'s being absent in standard differential renormalization. In
the next Section~6 we shall see another demonstration of their
importance.

The kinship of the EG method with differential renormalization
\textit{\`a la} Smirnov and Zavialov was recognized by
Prange~\cite{PrangeI}; he was stumped for nonlogarithmic divergences,
though. See~\cite{Cybele} in the same vein.

\subsection{The connection with BPHZ renormalization}

We still have left some chips to cash. We elaborate next the statement
that BPHZ subtraction has no independent status from Epstein--Glaser,
and that the validity of that renormalization method is just a
corollary of the latter. This involves just a two-line proof.

The Fourier transforms of the causally renormalized amplitudes exist
at least in the sense of tempered distributions. They are in fact
rather regular. Taking Fourier transforms is tantamount to replacing
the test function by an exponential, which, according to the Ces\`aro
theory of~\cite{Odysseus,CesarRicardo}, can preclude smoothness of the
momentum space amplitude only at the origin. The appearance of an
(integrable) singularity at $p = 0$ is physically expected in a theory
of massless particles.

Let us fix our conventions. We define Fourier and inverse Fourier
transforms by
$$
F[\phi](p) := \hat\phi(p)
:= \int \frac{d^dx}{(2\pi)^{d/2}}\, e^{-ipx} \phi(x),
$$
and
$$
F^{-1}[\phi](p) := \check\phi(p)
:= \int \frac{d^dx}{(2\pi)^{d/2}}\, e^{ipx} \phi(x),
$$
respectively. It follows that
$$
(x^\mu \phi)\check{\ }(p) = (-i)^\mu \del^\mu \check\phi(p),
$$
where $\mu$ denotes a multiindex; so that, in particular,
$$
(x^\mu)\check{\ }(p) = (-i)^\mu (2\pi)^{d/2} \dl^{(\mu)}(p).
$$
Also,
\begin{equation}
\del^\mu\phi(0) = (-i)^\mu (2\pi)^{-d/2} \<p^\mu, \check\phi>.
\label{eq:acabaramos}
\end{equation}
{}From this and the following consequence of~\eqref{eq:bas-def}:
$$
\<F[\tilde f], F^{-1}[\phi]> = \<F[f], F^{-1}[R^k_0\phi]>,
$$
there follows at once
$$
F[\tilde f](p) = R^k_0 F[f](p).
$$
This is nothing but the BPHZ subtraction rule in momentum space.

\smallskip

We hasten to add:

\begin{itemize}
\item
An expression such as $F[f]$ is \textit{not} a priori meaningless: it
is a well defined functional on the linear subspace of Schwartz
functions $\phi$ whose first momenta $\int p^\a\phi(p)\,d^d p$ up to
order $k+1$ happen to vanish. (This is the counterpart $F\SS_{k+1}$,
according to~\eqref{eq:acabaramos}, of the distributions on real space
acting on test functions vanishing up to order $k+1$ at the origin.)

\item
Moreover, explicit expressions for these functionals on the external
variables are given precisely by the unrenormalized momentum space
amplitudes!

This circumstance constitutes the (deceptive) advantage of the BPHZ
formalism for renormalization. We say deceptive because ---as
persuasively argued in~\cite{ThreefromRussia}--- the BPHZ method makes
no effective use of the recursive properties of renormalization (paper
III) and then, when using it, prodigious amounts of energy must go
into proving convergence of, and/or computing, the (rather horrendous)
resulting integrals, into showing that the Minkowskian counterparts
define \textit{bona fide} distributions\dots Much more natural to
remain on the nutritious ground of distribution theory on real space,
throughout. But this has never been done.

\item
Also, the $\del^\mu F[f](0)$ for $|\mu| \leq k$ exist for massive
theories.
\end{itemize}

For zero-mass models, the basic BPHZ scheme runs into trouble; this is
due naturally to the failure of $\del^\mu\hat f(0)$ to exist for
$|\mu|=k$, on account of the infrared problem. Now, one can perform
subtraction at some external momentum $q\neq0$, providing a mass
scale. This is just the Fourier-mirrored version of standard EG
renormalization, with weight function $e^{-iqx}$; one only has to
remember to use~\eqref{eq:bas-EGbis} instead of~\eqref{eq:bas-EG}.

It is patent, though, that this last subtraction is quite awkward in
practice, and will introduce in the Minkowskian context a
noncovariance which must be compensated by further subtractions. This
prompted Lowenstein and Zimmermann to introduce their ``soft mass
insertions''~\cite{LowZim}. Which amounts to an epicycle too many.

In the light of the approach advocated in this paper, there exist
several simpler and more physical strategies.

\begin{itemize}
\item
One strategy is to recruit our basic formula~\eqref{eq:imp-EG} in
momentum space
$$
F[f](p) - j^{k-1}_0 F[f](p) -
\sum_{|\mu|=k} \frac{\del^\mu F[fw](p)\,p^\mu}{\mu!}.
$$
Still with $w(x) = e^{-iqx}$, this leads at once to
$$
F[f](p) - j^{k-1}_0 F[f](p) -
\sum_{|\mu|=k} \frac{\del^\mu F[f](q)\,p^\mu}{\mu!}.
$$
Note that the difference between two of these recipes is polynomial in
$p^\mu$, with $|\mu|=k$ only, as it should. This can be more easily
corrected for Lorentz covariance, should the need
arise~\cite{Bettina}.

\item
A second method is to exploit homogeneity in adapting our recipes for
direct use in momentum space, in the spirit of~\cite{chavaloraro}
and~\cite{SmirIBP}.

\item
A third one is to perform Fourier analysis on our previous results.
One has
$$
\int \frac{d^4x}{(2\pi)^2}\, e^{-ipx}\, \frac{\log(\mu|x|)}{x^2} =
- \frac{1}{p^2} \log\biggl(\frac{C|p|}{2\mu}\biggr),
$$
where $C := e^\ga \simeq 1.781072\dots$ with $\ga$ the
Euler--Mascheroni constant. Then, from~\eqref{eq:nat-ren}, for
instance for the ``fish'' diagram in the $\varphi_4^4$ model:
$$
\widehat{[1/x^4]_R}(p) = \tquarter[1 - \log(C^2p^2/4\mu^2)],
$$
and more generally:
$$
\widehat{[1/x^{2k+4}]_R}(p) = \frac{(-)^{k+1}p^{2k}}{4^{k+1}k!(k+1)!}
\biggl[2\log\frac{|p|}{2\mu} - \Psi(k+1) - \Psi(k+2)\biggr],
$$
where $\Psi(x) := d/dx(\log\Ga(x))$ has been invoked, and we recall
that
$$
\Psi(n) = -\ga + H_{n-1}.
$$
For the setting sun diagram in the $\varphi_4^4$ model, in particular:
\begin{equation}
\widehat{[1/x^6]}_R =
\frac{p^2}{16} \biggl(\log\frac{|p|\ga}{2\mu} - \frac{5}{4}\biggr).
\label{eq:third-Fourier}
\end{equation}
\end{itemize}

\section{Some examples in massive theories}
\label{sec:massive-fun}

The aim of this short section is to dispel any idea that the
usefulness of EG-type renormalization, and in particular of the
$T$-subtraction, is restricted to massless models. The overall
conclusion, though, is that the massless theory keeps a normative
character. Our purposes being merely illustrative, we liberally borrow
from Schnetz~\cite{NR}, Prange~\cite{PrangeI} and Haagensen and
Latorre~\cite{HL}.

The first example is nothing short of spectacular. Suppose we add to
our original Lagrangian for $\varphi_4^4$ a mass term
$\thalf m^2\varphi^2$ and treat it as a perturbation, for the
calculation of the new propagator. Then we would have for $D_F(x)$:
$$
\frac{1}{x^2} - \int dx'\, \frac{m^2}{(x-x')^2 {x'}^2} +
\int dx'\,dx''\, \frac{m^4}{(x-x')^2 (x'-x'')^2 {x''}^2} - \cdots
$$
This ``nonrenormalizable'' interaction is tractable with our method.
We work in momentum space, so we just have to consider the
renormalization of $1/p^{2k}$ for $k > 1$. This is read directly
off~\eqref{eq:third-Fourier}, by inverting the roles of $p$ and~$x$,
with the proviso that $\mu$ gets replaced by $1/\mu$, in order to keep
the correct dimensions. Then the result is
$$
D_F(x) = \frac{1}{x^2} +
\frac{m^2}{2} \sum_{n=0}^\infty \frac{m^{2n}x^{2n}}{4^n\,n!(n+1)!}
\biggl(\log\frac{\mu|x|}{2} - \Psi(n+1) - \Psi(n+2)\biggr).
$$
On naturally identifying the scale $\mu = m$, one obtains on the nose
the \textit{exact} expansion of the \textit{exact} result
$$
D_F(x) = \frac{m}{|x|} K_1(m|x|).
$$
Here $K_1$ is the modified Bessel function of order~1. Had we kept the
original EG subtraction with a $H(\mu - |p|)$ weight, we would earn a
surfeit of terms with extra powers of $\mu$, landing in a serious
mess.

It is also interesting to see how well or badly fare the other
``competitors''. Differential renormalization gives a expression of
similar type but with different coefficients:
$$
\frac{1}{x^2} +
\frac{m^2}{2} \sum_{n=0}^\infty \frac{m^{2n}x^{2n}}{4^n\,n!(n+1)!}
\biggl( \log\frac{\mu|x|}{2} + 2\ga \biggr).
$$
To obtain the correct result, it is necessary to substitute a
different mass scale $\mu_n$ for each integral and to adjust ad hoc an
\textit{infinity} of such parameters. Dimensional regularization (plus
``minimal'' subtraction of a pole term for each summand but the first)
fares slightly better, as it ``only'' misses the $\Psi(n+2)$
terms~\cite{NR}. The distribution-theoretical rationale for the
success of the ``illegal'' expansion performed is explained
in~\cite{NR}.

Let us now look at the fish diagram in the massive theory. It is
possible to use~\eqref{eq:bas-Lagr} instead
of~\eqref{eq:Berkeley-jump}. Make the change of variables:
$$
t = \frac{|x|}{s};  \qquad  dt = -\frac{|x|\,ds}{s^2}.
$$
One gets, for the renormalized amplitude,
$$
-S\biggl[ \frac{m^2}{x^4} \int_{|x|}^\infty ds\, s K_1(ms)^2 \biggr].
$$
Now,
$$
\int ds\, s K_1(ms)^2 = \frac{s^2}{2} \biggl(
K_0^2(s) + 2 \frac{K_0(s)K_1(s)}{s} - K^2_1(s) \biggr)
$$
can be easily checked from
$$
K'_0(s) = K_1(s);  \qquad  K'_1(s) = - K_0(s) - \frac{K_1(s)}{s}.
$$
The final result is then
$$
\Dl\,\biggl[ \frac{m^2}{2} \Bigl( K^2_0(m|x|) - K_1^2(m|x|) +
\frac{K_0(m|x|)K_1(m|x|)}{m|x|} \Bigr) \biggr].
$$
Had we used~\eqref{eq:Berkeley-jump}, the upper limit of the integral
would become $1/\mu$, and the result would be modified by
$$
\frac{m^2}{4\mu^2x^2} (K^2_1(m/\mu) - K_2(m/\mu)K_0(m/\mu)).
$$
At the ``high energy'' limit, as $\mu \uparrow \infty$ and
$|x| \downarrow 0$, this interpolates between the previous result and
the renormalization in the massless case.

However, this method becomes cumbersome already for renormalizing
$D_F^3$. It is convenient to modify the strategy, and to use in this
context differential renormalization, corrected in such a way that the
known renormalized mass zero limit is kept. This idea succeeds because
of the good properties of our subtraction with respect to the mass
expansion. For instance, away from zero~\cite{HL},
$$
\biggl( \frac{mK_1(m|x|)}{|x|} \biggr)^3 =
\frac{m^2}{16} (\Dl - 9m^2)(\Dl - m^2) \bigl(
K_0(m|x|) K_1^2(m|x|) + K_0^3(m|x|) \bigr).
$$
Note the three-particle ``threshold''. To this Haagensen and Latorre
add a term of the form
$$
\frac{\pi^2}{4} \log\frac{2\mu}{\ga m} \,\Dl\dl(x),
$$
to which, for reasons sufficiently explained, we should add a term of
the form $\frac{5\pi^2}{8} \Dl\dl(x)$. A term proportional to~$\dl$
(thus a mass correction) is also present. As they indicate, it is
better fixed by a renormalization prescription.

\section{Conclusion}
\label{sec:le-clou}

We have delivered the missing link of the EG subtraction method to the
standard literature on extension of distributions. The improved
subtraction method sits at the crossroads in regard to dimensional
regularization in configuration space; differential renormalization;
``natural'' renormalization; and BPHZ renormalization. The discussions
in the previous sections go a long way to justify the conjecture (made
by Connes, and independently by Estrada) that Hadamard's finite part
theory is in principle enough to deal with quantum field theory
divergences. To accomplish that feat, however, it must go under the
guise of the $T$-projector; this gives the necessary flexibility to
deal with complicated diagrams with subdivergences~\cite{Flora}.

\bigskip

\subsection*{Acknowledgements}

I am indebted to Ph.~Blanchard for much encouragement and discussions,
to S.~Lazzarini for a computation pertaining to subsection~5.1, and to
D.~Kreimer and J.~C.~V\'arilly for comments on an earlier version of
the manuscript.

\end{document}